\documentclass[11pt,notitlepage]{article}%
\usepackage{amsfonts}
\usepackage{amsmath}
\usepackage{amssymb}
\usepackage{graphicx}%
\newtheorem{theorem}{Theorem}
\newtheorem{acknowledgement}[theorem]{Acknowledgement}

\newtheorem{corollary}[theorem]{Corollary}

\newtheorem{proposition}[theorem]{Proposition}

\newenvironment{proof}[1][Proof]{\noindent\textbf{#1.} }{\ \rule{0.5em}{0.5em}}
\begin{document}

\title{ON THE\ USEFULNESS OF\ MODULATION\ SPACES\ IN\ DEFORMATION\ QUANTIZATION}
\author{Maurice de Gosson \ and \ Franz Luef}
\maketitle

\begin{abstract}
We discuss the relevance to deformation quantization of Feichtinger's
modulation spaces, especially of the weighted Sj\"{o}strand classes
$M_{s}^{\infty,1}(\mathbb{R}^{2n})$. These function spaces are good classes of
symbols of pseudodifferential operators (observables). They have a widespread
use in time-frequency analysis and related topics, but are not very well-known
in physics. It turns out that they are particularly well adapted to the study
of the Moyal star-product and of the star-exponential.

\end{abstract}
\tableofcontents

\textbf{MSC\ (2000)}: 81S30, 81S10, 47G30


\section{Introduction}

It has become rather obvious since the 1990's that the theory of modulation
spaces, which plays a key role in time-frequency and Gabor analysis, often
allows to prove in a rather pedestrian way results that are usually studied
with methods of \textquotedblleft hard\textquotedblright\ analysis. These
spaces, whose definition goes back to the seminal work \cite{fe81,fei81,fe83}
of Feichtinger over the period 1980--1983 (also see Triebel \cite{Triebel})
are however not generally well-known by physicists, even those working in the
phase-space formulation of quantum mechanics. This is unfortunate, especially
since \textquotedblleft interdisciplinarity\textquotedblright\ has become so
fashionable in Science; it is a perfect example of two disciplines living in
mirror Universes, since, conversely, many techniques which have proven to be
successful in QM (for instance, symplectic geometry) are more or less ignored
in TFA (to be fair, Folland's book \cite{Folland} comes as close as possible
to such an interdisciplinary program, but this book was written in the 1980's,
and there has been much progress both in TFA and quantum mechanics since then).

This paper is a first (and modest) attempt towards the construction of bridges
between quantum mechanics in phase space, more precisely deformation
quantization, and these new and insufficiently exploited functional-analytic
techniques; this is made possible using the fact that ordinary (Weyl)
pseudo-differential calculus and deformation quantization are
\textquotedblleft intertwined\textquotedblright\ using the notion of
wave-packet transform, as we have shown in our recent paper \cite{golu08-2},
and the fact that these wave-packet transforms are closely related to the
windowed short-time Fourier transform appearing in the definition of
modulation spaces.

This work is structured as follows:

\begin{itemize}
\item In Section \ref{secone} we briefly review deformation quantization with
an emphasis on the point of view developed in de Gosson and Luef
\cite{golu08-2}; in this approach the star-product is expressed as the action
of a pseudo-differential operator $\widetilde{A}^{\hbar}$ of a certain type
(\textquotedblleft Bopp operator\textquotedblright). In fact, the Moyal
product $A\star_{\hbar}B$ of two observables can be expressed as
\begin{equation}
A\star_{\hbar}B=\widetilde{A}^{\hbar}(B)
\end{equation}
That operator is related to the usual Weyl operator by an intertwining formula
involving \textquotedblleft windowed wave-packet transforms\textquotedblright,
which are closely related to the short-time Fourier transform familiar from
time-frequency analysis. We take the opportunity to comment a recent statement
of Gerstenhaber on the choice of a \textquotedblright preferred
quantization\textquotedblright;

\item In Section \ref{secmod} we begin by introducing the basics of the theory
of modulation spaces we will need. We first introduce the weighted spaces
$M_{s}^{\infty,1}(\mathbb{R}^{2n})$ which generalize the so-called
Sj\"{o}strand classes. The elements of these spaces are very convenient as
pseudo-differential symbols (or \textquotedblleft
observables\textquotedblright); we show that, in particular, $M_{s}^{\infty
,1}(\mathbb{R}^{2n})$ is a $\ast$-algebra for the Moyal product (Proposition
\ref{propstar}): if $A,B\in M_{s}^{\infty,1}(\mathbb{R}^{2n})$ then
$A\star_{\hbar}B\in M_{s}^{\infty,1}(\mathbb{R}^{2n})$ and $\overline{A}\in
M_{s}^{\infty,1}(\mathbb{R}^{2n})$. We moreover prove the following
\textquotedblleft Wiener property\textquotedblright\ of the Moyal product: if
$A\in M_{s}^{\infty,1}(\mathbb{R}^{2n})$ and $A\star_{\hbar}B=I$ then $B\in
M_{s}^{\infty,1}(\mathbb{R}^{2n})$. We thereafter define the modulation spaces
$M_{s}^{q}(\mathbb{R}^{n})$ which are particularly convenient for describing
phase-space properties of wave-functions. The use of modulation spaces in
deformation quantization requires a redefinition of these spaces in terms of
the cross-Wigner transform. We do not consider here the slightly more general
spaces $M_{s}^{q,r}(\mathbb{R}^{n})$, this mainly for the sake of notational
brevity, however most of our results can be generalized without difficulty to
this case. We finally redefine the star-exponential
\begin{equation}
\operatorname{Exp}(Ht)=\sum_{k=0}^{\infty}\frac{1}{k!}\left(  \frac{t}{i\hbar
}\right)  ^{k}\widetilde{H}^{k};
\end{equation}
in terms of the Bopp operators; in fact we have%
\begin{equation}
\operatorname{Exp}(Ht)=\exp\left(  -\frac{i}{\hbar}\widetilde{H}t\right)  .
\end{equation}
This allows us to prove regularity results for $\operatorname{Exp}(Ht)$.
\end{itemize}

\subsection*{Notation}

The scalar product of two square integrable functions $\psi$ and $\psi
^{\prime}$ on $\mathbb{R}^{n}$ is written $(\psi|\psi^{\prime})$; that of
functions $\Psi,\Psi^{\prime}$ on $\mathbb{R}^{2n}$ is $((\Psi|\Psi^{\prime
}))$. We denote by $\mathcal{S}(\mathbb{R}^{n})$ the Schwartz space of
functions decreasing, together with their derivatives, faster than the inverse
of any polynomial. The dual $\mathcal{S}^{\prime}(\mathbb{R}^{n})$ of
$\mathcal{S}(\mathbb{R}^{n})$ is the space of tempered distributions. The
standard symplectic form on $\mathbb{R}^{n}\times\mathbb{R}^{n}\equiv
\mathbb{R}^{2n}$ is given by $\sigma(z,z^{\prime})=p\cdot x^{\prime}%
-p^{\prime}\cdot x$ if $z=(x,p)$ and $z^{\prime}=(x^{\prime},p^{\prime})$;
equivalently $\sigma(z,z^{\prime})=Jz\cdot z^{\prime}$ where $J=%
\begin{pmatrix}
0 & I\\
-I & 0
\end{pmatrix}
$ is the standard symplectic matrix. When using matrix notation $x,p,z$ are
viewed as column vectors.

If $A$ is a \textquotedblleft symbol\textquotedblright\ we denote
indifferently by $A^{w}(x,-i\hbar\partial_{x})$ or $\widehat{A}^{\hbar}$ the
corresponding Weyl operator.

We will also use multi-index notation: for $\alpha=(\alpha_{1},...,\alpha
_{2n})$ in $\mathbb{N}^{2n}$ we set%
\[
|\alpha|=\alpha_{1}+\cdot\cdot\cdot+\alpha_{2n}\text{ \ , \ }\partial
_{z}^{\alpha}=\partial_{z_{1}}^{\alpha_{1}}\cdot\cdot\cdot\partial_{z_{2n}%
}^{\alpha_{2n}}%
\]
where $\partial_{z_{j}}^{\alpha_{j}}=\partial^{\alpha_{j}}/\partial
x_{j}^{\alpha_{j}}$ for $1\leq j\leq n$ and $\partial_{z_{j}}^{\alpha_{j}%
}=\partial^{\alpha_{j}}/\partial\xi_{j}^{\alpha_{j}}$ for $n+1\leq j\leq2n$.

The unitary $\hbar$-Fourier transform is defined, for $\psi\in\mathcal{S}%
(\mathbb{R}^{n})$, by%
\[
F\psi(x)=\left(  \tfrac{1}{2\pi\hbar}\right)  ^{n/2}\int_{\mathbb{R}^{n}%
}e^{-\frac{i}{\hbar}x\cdot x^{\prime}}\psi(x^{\prime})dx^{\prime}.
\]

\section{Deformation Quantization and Bopp Calculus\label{secone}}

The rigorous definition of deformation quantization goes back to the work
\cite{BFFLS1,BFFLS2} of Bayen et al. in the end of the 1970s; also see the
contribution by Maillard \cite{maillard}. We recommend the reading of
Sternheimer's paper \cite{Stern1} for a recent discussion of the topic and its
genesis. The relation with a special Weyl calculus (which we call
\textquotedblleft Bopp calculus\textquotedblright) was introduced in de Gosson
and Luef

\subsection{Deformation quantization}

\subsubsection{Generalities}

Roughly speaking, the starting idea is that if we view classical mechanics as
the limit of quantum mechanics when $\hbar\rightarrow0$, then we should be
able to construct quantum mechanics by \textquotedblleft
deforming\textquotedblright\ classical mechanics. On the simplest level (which
is the one considered in this paper), one replaces the ordinary product of two
functions on phase space, say $A$ and $B$, by a \textquotedblleft star
product\textquotedblright%
\[
A\star_{\hbar}B=AB+\sum_{j=1}^{\infty}\hbar^{j}C_{j}(A,B)
\]
where the $C_{j}$ are certain bidifferential operators. Since one wants the
star-product to define an algebra structure, one imposes certain conditions on
$A\star_{\hbar}B$: it should be associative; moreover it should become the
ordinary product $AB$ in the limit $\hbar\rightarrow0$ and we should recover
the Poisson bracket $\{A,B\}$ from the quantity $i\hbar^{-1}(A\star_{\hbar
}B-B\star_{\hbar}A)$ when $\hbar\rightarrow0$.

Assume now that
\[
\widehat{A}^{\hbar}=A^{w}(x,-i\hbar\partial_{x}):\mathcal{S}(\mathbb{R}%
^{n})\longrightarrow\mathcal{S}^{\prime}(\mathbb{R}^{n})
\]
and
\[
\widehat{B}^{\hbar}=B^{w}(x,-i\hbar\partial_{x}):\mathcal{S}(\mathbb{R}%
^{n})\longrightarrow\mathcal{S}(\mathbb{R}^{n}).
\]
Then the product $\widehat{C}^{\hbar}=\widehat{A}^{\hbar}\widehat{B}^{\hbar}$
is defined on $\mathcal{S}(\mathbb{R}^{n})$ and we have $\widehat{C}^{\hbar
}=C^{w}(x,-i\hbar\partial_{x})$ where the symbol $C$ is given by the Moyal
product $C=A\star_{\hbar}B$:
\begin{equation}
A\star_{\hbar}B(z)=\left(  \tfrac{1}{4\pi\hbar}\right)  ^{2n}\iint
\nolimits_{\mathbb{R}^{2n}}e^{\frac{i}{2\hbar}\sigma(u,v)}A(z+\tfrac{1}%
{2}u)B(z-\tfrac{1}{2}v)\mathrm{d}u\mathrm{d}v \label{cz}%
\end{equation}
(historically formula (\ref{cz}) goes back to the seminal work of Moyal
\cite{Moyal} and Groenewold \cite{groen}). Equivalently:%
\begin{equation}
A\star_{\hbar}B(z)=\left(  \tfrac{1}{\pi\hbar}\right)  ^{2n}\iint
\nolimits_{\mathbb{R}^{2n}}e^{-\frac{2i}{\hbar}\sigma(z-z^{\prime}%
,z-z^{\prime\prime})}A(z^{\prime})B(z^{\prime\prime})\mathrm{d}z^{\prime
}\mathrm{d}z^{\prime\prime}. \label{star}%
\end{equation}
Recall that the Weyl symbol of an operator $\widehat{A}^{\hbar}:\mathcal{S}%
(\mathbb{R}^{n})\longrightarrow\mathcal{S}^{\prime}(\mathbb{R}^{n})$ is the
distribution $A\in\mathcal{S}^{\prime}(\mathbb{R}^{2n})$ such that
\begin{equation}
\widehat{A}^{\hbar}=\left(  \tfrac{1}{2\pi\hbar}\right)  ^{n}\int
_{\mathbb{R}^{2n}}A_{\sigma}(z_{0})\widehat{T}^{\hbar}(z_{0})\mathrm{d}z_{0}
\label{weyl}%
\end{equation}
where $\widehat{T}^{\hbar}(z_{0})$ is the Heisenberg--Weyl operator, defined
by%
\begin{equation}
\widehat{T}^{\hbar}(z_{0})\psi(x)=e^{\frac{i}{\hbar}(p_{0}\cdot x-\frac{1}%
{2}p_{0}\cdot x_{0})}\psi(x-x_{0}) \label{hw}%
\end{equation}
if $z_{0}=(x_{0},p_{0})$ and
\begin{equation}
A_{\sigma}(z)=F_{\sigma}A(z)=\left(  \tfrac{1}{2\pi\hbar}\right)  ^{n}%
\int_{\mathbb{R}^{2n}}e^{-\frac{i}{\hbar}\sigma(z,z^{\prime})}A(z^{\prime
})\mathrm{d}z \label{asa}%
\end{equation}
is the symplectic Fourier transform of $A$; note that $F_{\sigma}A(z)=FA(-Jz)$.

It is clear that the Moyal product is associative (because composition of
operators is); to see that $\lim_{\hbar\rightarrow0}A\star_{\hbar}B=AB$ it
suffices (at least on a formal level) to perform the change of variables
$(u,v)\longmapsto\sqrt{\hbar}(u,v)$ in the integral in (\ref{cz}), which leads
to%
\begin{equation}
A\star_{\hbar}B(z)=\left(  \tfrac{1}{4\pi}\right)  ^{2n}\iint
\nolimits_{\mathbb{R}^{2n}}e^{\frac{i}{2}\sigma(u,v)}A(z+\tfrac{\sqrt{\hbar}%
}{2}u)B(z-\tfrac{\sqrt{\hbar}}{2}v)\mathrm{d}u\mathrm{d}v; \label{abrep}%
\end{equation}
letting $\hbar\rightarrow0$ and using the Fourier inversion formula
\[
\iint\nolimits_{\mathbb{R}^{2n}}e^{\frac{i}{2}\sigma(u,v)}\mathrm{d}%
u\mathrm{d}v=\left(  4\pi\right)  ^{2n}.
\]
we get $\lim_{\hbar\rightarrow0}C(z)=A(z)B(z)$. That we also have
\[
\lim_{\hbar\rightarrow0}\left[  i\hbar^{-1}(A\star_{\hbar}B-B\star_{\hbar
}A)\right]  =\{A,B\}
\]
is verified in a similar way.

\subsubsection{Symplectic covariance}

Recall that the metaplectic group $\operatorname*{Mp}(2n,\mathbb{R})$ is the
unitary representation of the connected double covering of the symplectic
group $\operatorname*{Sp}(2n,\mathbb{R})$ (see e.g.
\cite{Folland,Birk,Littlejohn}). The metaplectic group is generated by the
following unitary operators:

\begin{itemize}
\item The modified $\hbar$-Fourier transform
\begin{equation}
\widehat{J}^{\hbar}=i^{-n/2}F \label{jh}%
\end{equation}
whose projection on $\operatorname*{Sp}(2n,\mathbb{R})$ is the standard
symplectic matrix $J$;

\item The \textquotedblleft chirps\textquotedblright\ $\widehat{V_{-P}}%
^{\hbar}$ defined, for $P=P^{T}$ by
\begin{equation}
\widehat{V_{-P}}^{\hbar}\psi(x)=e^{\frac{i}{\hbar}Px\cdot x}\psi(x) \label{vp}%
\end{equation}
whose projection on $\operatorname*{Sp}(2n,\mathbb{R})$ is $%
\begin{pmatrix}
I & 0\\
P & I
\end{pmatrix}
$;

\item The unitary changes of variables, defined for invertible $L$ by
\begin{equation}
\widehat{M_{L,m}}^{\hbar}\psi(x)=i^{m}\sqrt{|\det L|}\psi(Lx) \label{ml}%
\end{equation}
where the integer $m$ corresponds to a choice of $\arg\det L$; its projection
on $\operatorname*{Sp}(2n,\mathbb{R})$ is $%
\begin{pmatrix}
L^{-1} & 0\\
0 & L^{T}%
\end{pmatrix}
$.
\end{itemize}

Every $S\in\operatorname*{Sp}(2n,\mathbb{R})$ is the projection of two
operators $\pm\widehat{S}^{\hbar}$ in $\operatorname*{Mp}(2n,\mathbb{R})$.

We recall the following fundamental symplectic covariance property of Weyl
calculus:%
\begin{equation}
\widehat{(A\circ S^{-1})}^{\hbar}=\widehat{S}^{\hbar}\widehat{A}^{\hbar
}\widehat{S}^{\hbar-1} \label{syco1}%
\end{equation}
where $\widehat{S}^{\hbar}$ is any of the two metaplectic operators associated
with $S$.

\begin{proposition}
\label{sycobo}For every $S\in\operatorname*{Sp}(2n,\mathbb{R})$ we have%
\begin{equation}
(A\circ S^{-1})\star_{\hbar}=U_{S}(A\star_{\hbar})U_{S}^{-1} \label{coboppp}%
\end{equation}
where $U_{S}$ is the unitary operator on $L^{2}(\mathbb{R}^{2n})$ defined by
$U_{S}\Psi(z)=\Psi(Sz)$, and we have $U_{S}\in\operatorname*{Mp}%
(4n,\mathbb{R})$.
\end{proposition}

\begin{proof}
To prove (\ref{coboppp}) we notice that $A\star_{\hbar}$ is the Bopp operator
with Weyl symbol $\mathbb{A}(z,\zeta)=A(z-\tfrac{1}{2}J\zeta)$. Let
$\widetilde{A}^{S^{-1}}$ be the Weyl symbol of the Bopp operator
$\widetilde{H\circ S^{-1}}$; since $S^{-1}J=JSS^{T}$ we have%
\[
(\widetilde{A}^{\hbar})^{S^{-1}}(z,\zeta)=A(S^{-1}(z-\tfrac{1}{2}%
J\zeta))=\widetilde{A}^{\hbar}(M_{S}(z,\zeta))
\]
with
\begin{equation}
M_{S}=%
\begin{pmatrix}
S^{-1} & 0\\
0 & S^{T}%
\end{pmatrix}
\in\operatorname*{Sp}(4n,\mathbb{R}) \label{ms}%
\end{equation}
($\operatorname*{Sp}(4n,\mathbb{R})$ is the symplectic group of $\mathbb{R}%
^{4n}$ equipped with the standard symplectic form $\sigma\oplus\sigma$). It
follows from the general theory of the metaplectic group (see in particular
Proposition 7.8(i) in \cite{Birk}) that $M_{S}$ is the projection on
$\operatorname*{Sp}(4n,\mathbb{R})$ of the metaplectic operator $U_{S}$
defined by
\[
U_{S}\Psi(z)=\sqrt{\det S}\Psi(Sz)=\Psi(Sz)
\]
(recall that $\det S=1$). This proves (\ref{coboppp}) applying the covariance
formula (\ref{syco1}) to $\widetilde{H}$ viewed as a Weyl operator. That
$U_{S}\in\operatorname*{Mp}(4n,\mathbb{R})$ is clear (cf. formula (\ref{ml})).
\end{proof}

\subsubsection{On the use of Weyl calculus in deformation quantization}

We take the opportunity to briefly discuss a remark done by Gerstenhaber in
his recent paper \cite{ge07}. The Weyl correspondence resolves in a particular
way the ordering ambiguity when one passes from a symbol (\textquotedblleft
classical observable\textquotedblright) $A(x,p)$ to its quantized version
$A(\widehat{x},\widehat{p})$; for instance to monomials such as $xp$ or
$x^{2}p$ it associates the symmetrized operators $\frac{1}{2}(\widehat
{x}\widehat{p}+\widehat{p}\widehat{x})$ and $\frac{1}{3}(\widehat{x}%
^{2}\widehat{p}+\widehat{x}\widehat{p}+\widehat{p}\widehat{x}^{2})$. This
choice, argues Gerstenhaber, is totally arbitrary, and other choices are, a
priori, equally good (for instance, people working in partial differential
equations would usually choose the quantizations $\widehat{x}\widehat{p}$ and
$\widehat{x}^{2}\widehat{p}$ in the examples above), in fact for a given
symbol we have infinitely many choices
\begin{equation}
\widehat{A}_{\tau}^{\hbar}\psi(x)=\left(  \tfrac{1}{2\pi\hbar}\right)
^{n}\iint\nolimits_{\mathbb{R}^{2n}}e^{\frac{i}{\hbar}p\cdot(x-y)}%
A((1-\tau)x+\tau y,p)\psi(y)dydp \label{23.31}%
\end{equation}
corresponding to a parameter value $\tau$ (see Shubin \cite{sh87}); Weyl
quantization corresponds to the choice $\tau=1/2$. Gerstenhaber is right, no
doubt. However, one should understand that when working in deformation
quantization, the Weyl correspondence is still the most \textquotedblleft
natural\textquotedblright, and this for the following reason: the primary aim
of deformation quantization is to view quantum mechanics as a deformation of a
classical theory, namely classical mechanics in its Hamiltonian formulation.
Now, one of the main features of the Hamiltonian approach is its symplectic
covariance. It is therefore certainly desirable that the objects that one
introduces in a theory whose vocation is to mimic Hamiltonian mechanics
retains this fundamental feature. It turns out that not only is Weyl calculus
a symplectically covariant theory, but it is also the \textit{only}
quantization scheme having this property! This fact, which was already known
to Shale \cite{shale} (and is proven in detail in the last Chapter of Wong's
book \cite{Wong}) justifies a posteriori the suitability of the Weyl
correspondence in deformation quantization, as opposed to other ordering schemes.

\subsection{Moyal product and Bopp operators}

\subsubsection{The notion of Bopp pseudo-differential operator}

There is another way to write the Moyal product, which is reminiscent of
formula (\ref{weyl}) for Weyl pseudodifferential operators. Performing the
change of variables $v=z_{0}$, $z+\tfrac{1}{2}u=z^{\prime}$ in formula
(\ref{cz}) we get
\begin{align*}
A\star_{\hbar}B(z)  &  =\left(  \tfrac{1}{2\pi\hbar}\right)  ^{2n}%
\iint\nolimits_{\mathbb{R}^{4n}}e^{-\frac{i}{\hbar}\sigma(z_{0},z^{\prime}%
-z)}A(z^{\prime})B(z-\tfrac{1}{2}z_{0})\mathrm{d}z_{0}\mathrm{d}z^{\prime}\\
&  =\left(  \tfrac{1}{2\pi\hbar}\right)  ^{2n}\int_{\mathbb{R}^{2n}}\left[
\int_{\mathbb{R}^{2n}}e^{-\frac{i}{\hbar}\sigma(z_{0},z^{\prime})}A(z^{\prime
})\mathrm{d}z^{\prime}\right]  e^{\frac{i}{\hbar}\sigma(z_{0},z)}B(z-\tfrac
{1}{2}z_{0})\mathrm{d}z_{0}.
\end{align*}
Defining the operators $\widetilde{T}(z_{0}):\mathcal{S}(\mathbb{R}%
^{2n})\longrightarrow\mathcal{S}(\mathbb{R}^{2n})$ by%
\begin{equation}
\widetilde{T}(z_{0})B(z)=e^{\frac{i}{\hbar}\sigma(z_{0},z)}B(z-\tfrac{1}%
{2}z_{0}) \label{titi}%
\end{equation}
we can thus write the Moyal product in the form%
\begin{equation}
A\star_{\hbar}B=\left(  \tfrac{1}{2\pi\hbar}\right)  ^{n}\int_{\mathbb{R}%
^{2n}}A_{\sigma}(z_{0})(\widetilde{T}(z_{0})B)\mathrm{d}z_{0}. \label{ahb}%
\end{equation}
This formula, which is reminiscent of the representation (\ref{weyl}) of Weyl
operators, will play an important role in the subsequent sections. Note that
the operators $\widetilde{T}(z_{0})$ are unitary on $L^{2}(\mathbb{R}^{2n})$
and satisfy the same commutation relations as the Heisenberg--Weyl operators.

In \cite{golu08-2} we have proven the following results:

\begin{proposition}
The Weyl symbol of the operator
\begin{equation}
\widetilde{A}^{\hbar}:B\longmapsto\widetilde{A}^{\hbar}(B)=A\star_{\hbar}B
\label{ab}%
\end{equation}
is the distribution $\mathbb{A}\in\mathcal{S}^{\prime}(\mathbb{R}^{n}%
\times\mathbb{R}^{n})$ given by%
\begin{equation}
\mathbb{A}(z,\zeta)=A(z-\tfrac{1}{2}J\zeta)=A(x-\tfrac{1}{2}\zeta_{p}%
,p+\tfrac{1}{2}\zeta_{x}) \label{atild}%
\end{equation}
where $z\in\mathbb{R}^{2n}$ and $\zeta\in\mathbb{R}^{2n}$ are viewed as dual variables.
\end{proposition}

\subsubsection{Windowed wave-packet transforms}

For $\phi\in L^{2}(\mathbb{R}^{n})$ such that $||\phi||_{L^{2}}=1$ we define
the windowed wave-packet transform $W_{\phi}:\mathcal{S}^{\prime}%
(\mathbb{R}^{n})\longrightarrow\mathcal{S}(\mathbb{R}^{2n})$ by
\begin{equation}
W_{\phi}\psi=(2\pi\hbar)^{n/2}W(\psi,\phi) \label{wpt}%
\end{equation}
for $\psi\in\mathcal{S}^{\prime}(\mathbb{R}^{n})$; here $W(\psi,\phi)$ is the
usual cross-Wigner transform, given by%

\begin{equation}
W(\psi,\phi)(z)=\left(  \tfrac{1}{2\pi\hbar}\right)  ^{n}\int_{\mathbb{R}^{n}%
}e^{-\frac{i}{\hbar}p\cdot y}\psi(x+\tfrac{1}{2}y)\overline{\phi(x-\tfrac
{1}{2}y)}\mathrm{d}y. \label{wcr}%
\end{equation}
The windowed wave-packet transform is thus explicitly given by%
\[
W_{\phi}\psi(z)=\left(  \tfrac{1}{2\pi\hbar}\right)  ^{n/2}\int_{\mathbb{R}%
^{n}}e^{-\frac{i}{\hbar}p\cdot y}\psi(x+\tfrac{1}{2}y)\overline{\phi
(x-\tfrac{1}{2}y)}\mathrm{d}y.
\]
Since $||\phi||_{L^{2}}=1$ it follows from Moyal's identity
\begin{equation}
((W(\psi,\phi)|W(\psi^{\prime},\phi^{\prime})))=\left(  \tfrac{1}{2\pi\hbar
}\right)  ^{n}(\psi|\psi^{\prime})\overline{(\phi|\phi^{\prime})}
\label{moyal}%
\end{equation}
(see e.g. \cite{Birk,gr01}) that the restriction of $W_{\phi}$ to
$L^{2}(\mathbb{R}^{n})$ is a linear isometry of $L^{2}(\mathbb{R}^{n})$ onto a
subspace $\mathcal{H}_{\phi}$ of $L^{2}(\mathbb{R}^{2n})$. A simple
calculation shows that for $\Psi\in\mathcal{S}(\mathbb{R}^{n})$ the adjoint
$W_{\phi}^{\ast}:L^{2}(\mathbb{R}^{2n})\longrightarrow L^{2}(\mathbb{R}^{n})$
of $W_{\phi}$ is given by
\begin{equation}
W_{\phi}^{\ast}\Psi(x)=\left(  \tfrac{2}{\pi\hbar}\right)  ^{n/2}%
\int_{\mathbb{R}^{n}}e^{\frac{2i}{\hbar}p\cdot(x-y)}\phi(2y-x)\Psi
(y,p)\mathrm{d}p\mathrm{d}y. \label{uadj}%
\end{equation}
The subspace $\mathcal{H}_{\phi}$ is closed (and hence a Hilbert space): the
mapping $P_{\phi}=W_{\phi}W_{\phi}^{\ast}$ satisfies $P_{\phi}=P_{\phi}^{\ast
}$ and $P_{\phi}P_{\phi}^{\ast}=P_{\phi}$ hence $P_{\phi}$ is an orthogonal
projection. Since $W_{\phi}^{\ast}W_{\phi}$ is the identity on $L^{2}%
(\mathbb{R}^{n})$ the range of $W_{\phi}^{\ast}$ is $L^{2}(\mathbb{R}^{n})$
and that of $P_{\phi}$ is therefore precisely $\mathcal{H}_{\phi}$. Since
$\mathcal{H}_{\phi}$ is the range of $P_{\phi}$ and the closedness of
$\mathcal{H}_{\phi}$ follows.

\subsubsection{The intertwining property}

The key to the relation between deformation quantization and Bopp calculus
comes from following result:

\begin{proposition}
We have the intertwining formulae%
\begin{equation}
\widetilde{A}^{\hbar}W_{\phi}=W_{\phi}\widehat{A}^{\hbar}\text{ \ , \ }%
W_{\phi}^{\ast}\widetilde{A}^{\hbar}=\widehat{A}^{\hbar}W_{\phi}^{\ast}
\label{fund}%
\end{equation}
where $W_{\phi}^{\ast}:\mathcal{S}(\mathbb{R}^{2n})\longrightarrow
\mathcal{S}(\mathbb{R}^{n})$ is the adjoint of $W_{\phi}$. Equivalently:%
\begin{equation}
A\star_{\hbar}(W_{\phi}\psi)=W_{\phi}(\widehat{A}^{\hbar}\psi)\text{ \ ,
\ }W_{\phi}^{\ast}(A\star_{\hbar}B)=\widehat{A}^{\hbar}(W_{\phi}^{\ast}B)
\label{fundstar}%
\end{equation}
for $\psi\in\mathcal{S}(\mathbb{R}^{n})$.
\end{proposition}

\begin{proof}
See Proposition 2 in \cite{golu08-2}.
\end{proof}

Formula (\ref{atild}) justifies the notation%
\begin{equation}
\widetilde{A}^{\hbar}=A(x+\tfrac{1}{2}i\hbar\partial_{p},p-\tfrac{1}{2}%
i\hbar\partial_{x}) \label{ati}%
\end{equation}
and we will call $\widetilde{A}^{\hbar}$ the \emph{Bopp pseudo-differential
operator with symbol} $A$; the terminology is inspired by the paper
\cite{bopp} by Bopp, who was apparently the first to suggest the use of the
non-standard quantization rules%
\begin{equation}
(x,p)\longmapsto(x+\tfrac{1}{2}i\hbar\partial_{p},p-\tfrac{1}{2}i\hbar
\partial_{x}) \label{qr}%
\end{equation}
(which also appear in Kubo's paper \cite{kubo}). We note that formula
(\ref{ati}) is found in many physical texts \textit{without justification}. It
was precisely one of the aims of \cite{golu08-2} to give a rigorous
justification of this notation.

\section{Modulation Spaces\label{secmod}}

We define and list the main properties of two particular types of modulation
spaces: the spaces $M_{s}^{\infty,1}(\mathbb{R}^{2n})$ which are a
generalization of the Sj\"{o}strand classes, and the spaces $M_{s}%
^{q}(\mathbb{R}^{n})$ which contain, as a particular case the Feichtinger
algebra. We refer to Gr\"{o}chenig's book \cite{gr01} for proofs and generalizations.

\subsection{A good symbol (=observable) class: $M_{s}^{\infty,1}%
(\mathbb{R}^{2n})$}

In \cite{sj94,sj95} Sj\"{o}strand introduced a class of symbols which was
shown by Gr\"{o}chenig \cite{gr06bis} to be identical with the modulation
space%
\[
M_{0}^{\infty,1}(\mathbb{R}^{2n})=M^{\infty,1}(\mathbb{R}^{2n}).
\]
The Sj\"{o}strand class $M^{\infty,1}(\mathbb{R}^{2n})$ contains, in
particular, the symbol class $S_{0,0}^{0}(\mathbb{R}^{2n})$ consisting of all
infinitely differentiable complex functions $A$ on $\mathbb{R}^{2n}$ such that
$\partial_{z}^{\alpha}A$ is bounded for all multi-indices $\alpha\in
\mathbb{N}^{2n}$.

In this section we study a weighted version of the Sj\"{o}strand class.

\subsubsection{Definition and main properties}

In the 1970's the study of $L^{2}$-boundedness of pseudo-differential
operators was a popular area of research. For instance, a landmark was the
proof by Calder\'{o}n and Vaillancourt \cite{cv} that every operator with
symbol in $C^{2n+1}(\mathbb{R}^{2n})$ satisfying an additional condition had
this property (the same applies to the H\"{o}rmander class $S_{0,0}%
^{0}(\mathbb{R}^{2n})$). It turns out that results of this type --whose proofs
needed methods from hard analysis-- are much better understood (and easier
proved) using the theory of modulation spaces. For instance, Calder\'{o}n and
Vaillancourt's theorem is a simple corollary of the theory of the modulation
space of this subsection.

Let us introduce the weight function $v_{s}$ on $\mathbb{R}^{2n}$, defined for
$s\geq0$, by%
\begin{equation}
v_{s}(z,\zeta)=(1+|z|^{2}+|\zeta|^{2})^{s/2} \label{v2n}%
\end{equation}
(some of the results we list below remain valid for more general weight
functions). By definition, $M_{s}^{\infty,1}(\mathbb{R}^{2n})$ consists of all
$A\in\mathcal{S}^{\prime}(\mathbb{R}^{2n})$ such that there exists a function
$\Phi\in\mathcal{S}(\mathbb{R}^{2n})$ for which%
\begin{equation}
\sup_{z\in\mathbb{R}^{2n}}\left[  |V_{\Phi}A(z,\cdot)|v_{s}(z,\cdot)\right]
\in L^{1}(\mathbb{R}^{2n}) \label{supv}%
\end{equation}
where $V_{\Phi}A$ is the short-time Fourier transform of $A$ windowed by
$\Phi$:%
\begin{equation}
V_{\Phi}A(z,\zeta)=\int_{\mathbb{R}^{2n}}e^{-2\pi i\zeta\cdot z^{\prime}%
}A(z^{\prime})\overline{\Phi(z^{\prime}-z)}\mathrm{d}z^{\prime}. \label{stft2}%
\end{equation}
In particular, the Sj\"{o}strand class $M^{\infty,1}(\mathbb{R}^{2n})$ thus
consists of all $A\in\mathcal{S}^{\prime}(\mathbb{R}^{2n})$ such that%
\[
\int_{\mathbb{R}^{2n}}\sup_{z\in\mathbb{R}^{2n}}|V_{\Phi}A(z,\zeta
)|\mathrm{d}\zeta<\infty
\]
for some window $\Phi$. The formula
\begin{equation}
||A||_{M_{s}^{\infty,1}}^{\Phi}=\int_{\mathbb{R}^{2n}}\sup_{z\in
\mathbb{R}^{2n}}\left[  |V_{\Phi}A(z,\zeta)|v_{s}(z,\zeta)\right]
\mathrm{d}\zeta<\infty\label{modinf}%
\end{equation}
defines a norm on $M_{s}^{\infty,1}(\mathbb{R}^{2n})$. A remarkable (and
certainly not immediately obvious!) fact is that if condition (\ref{modinf})
holds for one window $\Phi$, then it holds for all windows; moreover when
$\Phi$ runs through $\mathcal{S}(\mathbb{R}^{2n})$ the functions
$||\cdot||_{M_{s}^{\infty,1}}^{\Phi}$ form a family of equivalent norms on
$M_{s}^{\infty,1}(\mathbb{R}^{2n})$. It turns out that $M_{s}^{\infty
,1}(\mathbb{R}^{2n})$ is a Banach space for the topology defined by any of
these norms; moreover the Schwartz space $\mathcal{S}(\mathbb{R}^{2n})$ is
dense in $M_{s}^{\infty,1}(\mathbb{R}^{2n})$.

The spaces $M_{s}^{\infty,1}(\mathbb{R}^{2n})$ are invariant under linear
changes of variables:

\begin{proposition}
\label{proplinv}Let $M$ be a real invertible $2n\times2n$ matrix. If $A\in
M_{s}^{\infty,1}(\mathbb{R}^{2n})$ then $A\circ M\in M_{s}^{\infty
,1}(\mathbb{R}^{2n}).$ In fact, there exists a constant $C_{M}>0$ such that
for every window $\Phi$ and every $A\in M_{s}^{\infty,1}(\mathbb{R}^{2n})$ we
have%
\begin{equation}
||A\circ M||_{M_{s}^{\infty,1}}^{\Phi}\leq C_{M}||A||_{M_{s}^{\infty,1}}%
^{\Psi} \label{estab}%
\end{equation}
where $\Psi=\Phi\circ M^{-1}$.
\end{proposition}

\begin{proof}
it suffices to prove the estimate (\ref{estab}) since $A\in M_{s}^{\infty
,1}(\mathbb{R}^{2n})$ if and only if $||A||_{M_{s}^{\infty,1}}^{\Psi}<\infty$.
Let us set $B=A\circ M$; performing the change of variables $z^{\prime
}\longmapsto Mz^{\prime}$ we have%
\[
V_{\Phi}B(z,\zeta)=(\det M)^{-1}\int_{\mathbb{R}^{2n}}e^{-2\pi i\zeta\cdot
M^{-1}z^{\prime}}A(z^{\prime})\overline{\Phi(M^{-1}z^{\prime}-z)}%
\mathrm{d}z^{\prime}%
\]
and hence%
\[
V_{\Phi}B(M^{-1}z,M^{T}\zeta)=(\det M)^{-1}\int_{\mathbb{R}^{2n}}e^{-2\pi
i\zeta\cdot z^{\prime}}A(z^{\prime})\overline{\Phi(M^{-1}(z^{\prime}%
-z))}\mathrm{d}z^{\prime}%
\]
that is%
\[
V_{\Phi}B(z,\zeta)=(\det M)^{-1}V_{\Psi}A(Mz,(M^{T})^{-1}\zeta)\text{ \ ,
}\Psi=\Phi\circ M^{-1}.
\]
It follows that%
\[
\sup_{z\in\mathbb{R}^{2n}}\left[  |V_{\Phi}B(z,\zeta)|v_{s}(z,\zeta)\right]
=(\det M)^{-1}\sup_{z\in\mathbb{R}^{2n}}\left[  V_{\Psi}A(z,(M^{T})^{-1}%
\zeta)v_{s}(M^{-1}z,\zeta)\right]
\]
so that%
\begin{align*}
||B||_{M_{s}^{\infty,1}}^{\Phi}  &  =(\det M)^{-1}\int_{\mathbb{R}^{2n}}%
\sup_{z\in\mathbb{R}^{2n}}\left[  V_{\Psi}A(z,(M^{T})^{-1}\zeta)v_{s}%
(M^{-1}z,\zeta)\right]  \mathrm{d}\zeta\\
&  =\int_{\mathbb{R}^{2n}}\sup_{z\in\mathbb{R}^{2n}}\left[  V_{\Psi}%
A(z,\zeta)v_{s}(M^{-1}z,M^{T}\zeta)\right]  \mathrm{d}\zeta.
\end{align*}
Diagonalizing $M$ and using the rotational invariance of $v_{s}$ it is easy to
see that there exists a constant $C_{M}$ such that%
\[
v_{s}(M^{-1}z,M^{T}\zeta)\leq C_{M}v_{s}(z,\zeta)
\]
and hence the inequality (\ref{estab}).
\end{proof}

The modulation spaces $M_{s}^{\infty,1}(\mathbb{R}^{2n})$ contain many of the
usual pseudo-differential symbol classes and we have the inclusion%
\begin{equation}
C_{b}^{2n+1}(\mathbb{R}^{2n})\subset M_{0}^{\infty,1}(\mathbb{R}^{2n})
\label{cm}%
\end{equation}
where $C_{b}^{2n+1}(\mathbb{R}^{2n})$ is the vector space of all functions
which are differentiable up to order $2n+1$ with bounded derivatives. In fact,
for every window $\Phi$ there exists a constant $C_{\Phi}>0$ such that%
\[
||A||_{M_{s}^{\infty,1}}^{\Phi}\leq C_{\Phi}||A||_{C^{2n+1}}=C_{\Phi}%
\sum_{|\alpha|\leq2n+1}||\partial_{z}^{\alpha}A||_{\infty}.
\]

\subsubsection{The $\ast$-algebra and Wiener properties}

Recall that the twisted product $A\#B$ of two symbols is defined by the
formula%
\begin{equation}
A\#B(z)=4^{n}\iint\nolimits_{\mathbb{R}^{2n}}e^{-4\pi i\sigma(z-z^{\prime
},z-z^{\prime\prime})}A(z^{\prime})B(z^{\prime\prime})\mathrm{d}z^{\prime
}\mathrm{d}z^{\prime\prime}. \label{abtw}%
\end{equation}

For us the main interest of $M_{s}^{\infty,1}(\mathbb{R}^{2n})$ comes from the
following property of the twisted product (Gr\"{o}chenig \cite{gr06bis}):

\begin{proposition}
Let $A,B\in M_{s}^{\infty,1}(\mathbb{R}^{2n})$. Then $A\#B\in M_{s}^{\infty
,1}(\mathbb{R}^{2n})$. In particular, for every window $\Phi$ there exists a
constant $C_{\Phi}>0$ such that
\[
||A\#B||_{M_{s}^{\infty,1}}^{\Phi}\leq C_{\Phi}||A||_{M_{s}^{\infty,1}}^{\Phi
}||B||_{M_{s}^{\infty,1}}^{\Phi}.
\]
Since obviously $\overline{A}\in M_{s}^{\infty,1}(\mathbb{R}^{2n})$ if and
only and $A\in M_{s}^{\infty,1}(\mathbb{R}^{2n})$ the property above can be
restated as:
\end{proposition}

\begin{quotation}
\textit{The modulation space }$M_{s}^{\infty,1}(\mathbb{R}^{2n})$\textit{ is a
Banach }$\ast$\textit{-algebra with respect to the twisted product }%
$\#$\textit{ and the involution }$A\longmapsto\overline{A}$\textit{. }
\end{quotation}

The Sj\"{o}strand class $M^{\infty,1}(\mathbb{R}^{2n})$ has the following property:

\begin{proposition}
\label{propsjo}(i) Every Weyl operator $\widehat{A}^{2\pi}$ with $A\in
M^{\infty,1}(\mathbb{R}^{2n})$ is bounded on $L^{2}(\mathbb{R}^{n})$; (ii) If
we have $\widehat{C}^{2\pi}=\widehat{A}^{2\pi}\widehat{B}^{2\pi}$with $A,B\in
M^{\infty,1}(\mathbb{R}^{2n})$ then $C\in M^{\infty,1}(\mathbb{R}^{2n})$;
(iii) If $\widehat{A}^{2\pi}$ with $A\in M^{\infty,1}(\mathbb{R}^{2n})$ is
invertible in $L^{2}(\mathbb{R}^{n})$ with inverse $\widehat{B}^{2\pi}$ then
$B\in M^{\infty,1}(\mathbb{R}^{2n})$.
\end{proposition}

Property (i) thus extends the $L^{2}$-boundedness property of operators with
symbols in $S_{0,0}^{0}(\mathbb{R}^{2n})$. Property (iii) is called the
\textit{Wiener property} of $M^{\infty,1}(\mathbb{R}^{2n})$; for the classical
symbol classes results of this type go back to Beals \cite{beals}; in the
recent paper by {Gr\"{o}chenig} and Rzeszotnik \cite{grrzXX} Beal's results
are proven using Banach space methods.

In \cite{gr06bis} Gr\"{o}chenig extended the Wiener property to the weighted
case. In fact, it follows from Theorem 4.6 (ibid.) that:

\begin{proposition}
\label{propsjogr} If $A\in M_{s}^{\infty,1}(\mathbb{R}^{2n})$ and $\widehat
{A}^{2\pi}$ is invertible on $L^{2}(\mathbb{R}^{n})$ then $(\widehat{A}^{2\pi
})^{-1}=\widehat{B}^{2\pi}$ with $B\in M^{\infty,1}(\mathbb{R}^{2n})$.
\end{proposition}

We will apply this important result to deformation quantization in Subsection
\ref{subsapp}.

\subsection{The modulation spaces $M_{s}^{q}(\mathbb{R}^{n})$}

\subsubsection{Definitions}

We define a weight $v_{s}$ on $\mathbb{R}^{2n}$ by
\begin{equation}
v_{s}(z)=(1+|z|^{2})^{s/2} \label{vs}%
\end{equation}
(cf. (\ref{v2n})). Notice that $v_{s}$ is submultiplicative:%
\begin{equation}
v_{s}(z+z^{\prime})\leq v_{s}(z)v_{s}(z^{\prime}). \label{submul}%
\end{equation}

In what follows $q$ is a real number $\geq1$, or $\infty$. Let $L_{s}%
^{q}(\mathbb{R}^{2n})$ be the space of all Lebesgue-measurable functions
$\Psi$ on $\mathbb{R}^{2n}$ such that $v_{s}\Psi\in L_{s}^{q}(\mathbb{R}%
^{2n})$. When $q<\infty$ the formula
\[
||\Psi||_{L_{s}^{q}}=\left(  \int_{\mathbb{R}^{2n}}|v_{s}(z)\Psi
(z)|^{q}\mathrm{d}z\right)  ^{1/q}%
\]
defines a norm on $L_{s}^{q}(\mathbb{R}^{2n})$; in the case $q=\infty$ this
formula is replaced by
\[
||\Psi||_{L_{s}^{\infty}}=\underset{z\in\mathbb{R}^{2n}}{\operatorname*{ess}%
\sup}|v_{s}(z)\Psi(z)|.
\]
The modulation space $M_{s}^{q}(\mathbb{R}^{n})$ is the vector space
consisting of all $\psi\in\mathcal{S}^{\prime}(\mathbb{R}^{n})$ such that
$V_{\phi}\psi\in L_{s}^{q}(\mathbb{R}^{2n})$ where $V_{\phi}$ is the
short-time Fourier transform (STFT) with window\textbf{ }$\phi\in
\mathcal{S}(\mathbb{R}^{n})$:%
\begin{equation}
V_{\phi}\psi(z)=\int_{\mathbb{R}^{n}}e^{-2\pi ip\cdot x^{\prime}}%
\psi(x^{\prime})\overline{\phi(x^{\prime}-x)}\mathrm{d}x^{\prime};
\label{stft}%
\end{equation}
it is related to the wave-packet transform by the formula%
\begin{equation}
W_{\phi}\psi(z)=2^{n}e^{\frac{2i}{\hbar}p\cdot x}V_{\phi_{\sqrt{2\pi\hbar}%
}^{\vee}}\psi_{\sqrt{2\pi\hbar}}(\sqrt{\tfrac{2}{\pi\hbar}}z) \label{wspt}%
\end{equation}
where $\phi^{\vee}(x)=\phi(-x)$ and $\psi_{\sqrt{2\pi\hbar}}(x)=\psi
(x\sqrt{2\pi\hbar})$. We thus have $\psi\in M_{s}^{q}(\mathbb{R}^{n})$ if and
only if there exists $\phi\in\mathcal{S}(\mathbb{R}^{n})$ such that
\begin{equation}
||\psi||_{M_{s}^{q}}^{\phi}=\left(  \int_{\mathbb{R}^{2n}}|v_{s}(z)V_{\phi
}\psi(z)|^{q}\mathrm{d}z\right)  ^{1/q}<\infty\label{mod2}%
\end{equation}
when $q<\infty$, and%
\begin{equation}
||\psi||_{M_{s}^{\infty}}^{\phi}=\underset{z\in\mathbb{R}^{2n}}%
{\operatorname*{ess}\sup}|v_{s}(z)V_{\phi}\psi(z)|<\infty\label{mod3}%
\end{equation}
when $q=\infty$. As in the case of the spaces $M_{s}^{\infty,1}(\mathbb{R}%
^{2n})$ this definition is independent of the choice of the \textquotedblleft
window\textquotedblright\ $\phi$, and the $||\cdot||_{M_{s}^{q}}^{\phi}$ form
a family of equivalent norms on $M_{s}^{q}(\mathbb{R}^{n})$, which is a Banach
space for the topology thus defined (see \cite{gr01}, Proposition 11.3.2,
p.233). Moreover the Schwartz space $\mathcal{S}(\mathbb{R}^{n})$ is dense in
$M_{s}^{q}(\mathbb{R}^{n})$.

The modulation spaces $M_{s}^{q}(\mathbb{R}^{n})$ can be redefined in terms of
the windowed wave-packet transform.

\begin{proposition}
\label{wips}We have $\psi\in M_{s}^{q}(\mathbb{R}^{n})$ if and only if
$W_{\phi}\psi\in L_{s}^{q}(\mathbb{R}^{2n})$ for some (and hence all) $\phi
\in\mathcal{S}(\mathbb{R}^{n})$.
\end{proposition}

\begin{proof}
It is based on formula (\ref{wspt}) relating the STFT $V_{\phi}$ to the
windowed wave-packet transform $W_{\phi}$. We only give the proof in the case
$1\leq q<\infty$, because the modifications needed in the case $q=\infty$ are
obvious. We have $\psi\in M_{s}^{q}(\mathbb{R}^{n})$ if and only if $V_{\phi
}\psi\in L_{s}^{q}(\mathbb{R}^{2n})$ for one (and hence every) $\phi
\in\mathcal{S}(\mathbb{R}^{n})$, that is if and only if $V_{\phi_{\sqrt
{2\pi\hbar}}^{\vee}}\psi\in L_{s}^{q}(\mathbb{R}^{2n})$. Since, in addition,
$\psi\in M_{s}^{q}(\mathbb{R}^{n})$ if and only if $\psi_{\sqrt{2\pi\hbar}}\in
M_{s}^{q}(\mathbb{R}^{n})$, we thus have $\psi\in M_{s}^{q}(\mathbb{R}^{n})$
if and only if $V_{\phi_{\sqrt{2\pi\hbar}}^{\vee}}\psi_{\sqrt{2\pi\hbar}}\in
L_{s}^{q}(\mathbb{R}^{2n})$ or, which amounts to the same,%
\begin{equation}
\psi\in M_{s}^{q}(\mathbb{R}^{n})\Longleftrightarrow2^{n}e^{\frac{2i}{\hbar
}p\cdot x}V_{\phi_{\sqrt{2\pi\hbar}}^{\vee}}\psi_{\sqrt{2\pi\hbar}}\in
L_{s}^{q}(\mathbb{R}^{2n}). \label{equiv}%
\end{equation}
(Recall that we denote $\psi_{\lambda}$ the function defined by $\psi
_{\lambda}(x)=\psi(\lambda x)$.) Now, a function $\Psi$ is in $L_{s}%
^{q}(\mathbb{R}^{2n})$ if and only if $\Psi_{\lambda}$ is, as follows from the
inequality%
\[
\int_{\mathbb{R}^{2n}}|v_{s}(z)\Psi(\lambda z)|^{q}\mathrm{d}z\leq
\lambda^{-2nq}(1+\lambda^{-2})^{s/2}\int_{\mathbb{R}^{n}}|v_{s}(z)\Psi
(z)|^{q}\mathrm{d}z
\]
\ obtained by performing the change of variable $z\longmapsto\lambda^{-1}z$
and the trivial estimate%
\[
(1+|\lambda^{-1}z|^{2})^{s/2}\leq(1+\lambda^{-2})^{s/2}(1+|z|^{2})^{s/2}%
\]
valid for all $s\geq0$. Combining this property (with $\lambda=\sqrt
{2/\pi\hbar}$) with the equivalence (\ref{equiv}), and using (\ref{wspt}), we
thus have $\psi\in M_{s}^{q}(\mathbb{R}^{n})$ if and only if $W_{\phi}\psi\in
L_{s}^{q}(\mathbb{R}^{2n})$.
\end{proof}

\subsubsection{Metaplectic and Heisenberg--Weyl invariance properties}

The modulation spaces $M_{s}^{q}(\mathbb{R}^{n})$ have the two remarkable
invariance properties.

\begin{proposition}
\label{propinvariance}(i) Each space $M_{s}^{q}(\mathbb{R}^{n})$ is invariant
under the action of the Heisenberg--Weyl operators $\widehat{T}^{\hbar}(z)$;
in fact there exists a constant $C>0$ such that%
\begin{equation}
||\widehat{T}^{\hbar}(z)\psi||_{M_{s}^{q}}^{\phi}\leq Cv_{s}(z)||\psi
||_{M_{s}^{q}}^{\phi}. \label{cv}%
\end{equation}
(ii) For $1\leq q<\infty$ the space $M_{s}^{q}(\mathbb{R}^{n})$ is invariant
under the action of the metaplectic group $\operatorname*{Mp}(2n,\mathbb{R})$:
if $\widehat{S}^{\hbar}\in\operatorname*{Mp}(2n,\mathbb{R})$ then $\widehat
{S}^{\hbar}\psi\in M_{s}^{q}(\mathbb{R}^{n})$ if and only if $\psi\in
M_{s}^{q}(\mathbb{R}^{n})$. In particular $M_{s}^{q}(\mathbb{R}^{n})$ is
invariant under the Fourier transform.
\end{proposition}

\begin{proof}
(i) The cross-Wigner transform satisfies%
\begin{align*}
W(\widehat{T}^{\hbar}(z_{0})\psi,\phi)(z)  &  =T(z_{0})W(\psi,\phi)(z)\\
&  =W(\psi,\phi)(z-z_{0})
\end{align*}
hence it suffices in view of Proposition \ref{wips} and definition (\ref{wpt})
to show that $L_{s}^{q}(\mathbb{R}^{2n})$ is invariant under the phase space
translation $T(z_{0})$. In view of the submultiplicative property
(\ref{submul}) of the weight $v_{s}$ we have, for $q<\infty$,%
\begin{align*}
||T(z_{0})\Psi||_{L_{v}^{q}}^{q}  &  =\int_{\mathbb{R}^{2n}}|\Psi
(z-z_{0})|^{q}v_{s}(z)^{q}dz\\
&  =\int_{\mathbb{R}^{2n}}|\Psi(z)|^{q}v_{s}(z+z_{0})^{q}dz\\
&  \leq v(z_{0})\int_{\mathbb{R}^{2n}}|\Psi(z)|^{q}v_{s}(z)^{q}dz
\end{align*}
hence our claim; the estimate (\ref{cv}) follows. A similar argument works in
the case $q=\infty$.
\end{proof}

The following consequence of the result above is the analogue of Proposition
\ref{proplinv}:

\begin{corollary}
\label{lemmah}The modulation space $M_{s}^{q}(\mathbb{R}^{n})$ is invariant
under the rescalings $\psi\longmapsto\psi_{\lambda}$ where $\psi_{\lambda
}(x)=\psi(\lambda x)$ where $\lambda\neq0$. More generally, $M_{s}%
^{q}(\mathbb{R}^{n})$ is invariant under every change of variables
$x\longmapsto Lx$ ($\det L\neq0$).
\end{corollary}

\begin{proof}
The unitary operators $M_{L}$ with $M_{L,m}\psi(x)=i^{m}\sqrt{|\det L|}%
\psi(Lx)$ ($\det L\neq0$, $\arg\det L\equiv m\pi$ $\operatorname{mod}2\pi$)
belong to $\operatorname*{Mp}(2n,\mathbb{R})$; the Lemma follows since
$M_{s}^{q}(\mathbb{R}^{n})$ is a vector space.
\end{proof}

The class of modulation spaces $M_{s}^{q}(\mathbb{R}^{n})$ contain as
particular cases many of the classical function spaces. For instance,
$M_{s}^{2}(\mathbb{R}^{n})$ coincides with the Shubin--Sobolev space
\[
Q^{s}(\mathbb{R}^{2n})=L_{s}^{2}(\mathbb{R}^{n})\cap H^{s}(\mathbb{R}^{n})
\]
(Shubin \cite{sh87}, p.45). We also have
\[
\mathcal{S}(\mathbb{R}^{n})=\bigcap_{s\geq0}M_{s}^{2}(\mathbb{R}^{n}).
\]

\subsubsection{The Feichtinger algebra}

A particularly interesting example of modulation space is obtained by taking
$q=1$ and $s=0$; the corresponding space $M_{0}^{1}(\mathbb{R}^{n})$ is often
denoted by $S_{0}(\mathbb{R}^{n})$, and is called the \textit{Feichtinger
algebra} \cite{fe83} (it is an algebra both for pointwise product and for
convolution). We have the inclusions%
\begin{equation}
\mathcal{S}(\mathbb{R}^{n})\subset S_{0}(\mathbb{R}^{n})\subset C^{0}%
(\mathbb{R}^{n})\cap L^{1}(\mathbb{R}^{n})\cap L^{2}(\mathbb{R}^{n}).
\label{incl}%
\end{equation}
A remarkable property of the Feichtinger algebra is that is the smallest
Banach space invariant under the action of the Heisenberg--Weyl operators
(\ref{hw}):

\begin{proposition}
Let $(\mathcal{B},||\cdot||)$ be a Banach algebra of tempered distributions on
$\mathbb{R}^{n}$. Suppose that $\mathcal{B}$ satisfies the two following
conditions: (i) there exists $C>0$ such that
\[
||\widehat{T}^{\hbar}(z)\psi||\leq Cv_{s}(z)||\psi||
\]
for all $z\in\mathbb{R}^{2n}$ and $\psi\in\mathcal{B}$; (ii) $M_{s}%
^{1}(\mathbb{R}^{n})\cap\mathcal{B}\neq\{0\}$. Then $M_{s}^{1}(\mathbb{R}%
^{n})$ is embedded in $\mathcal{B}$ and $S_{0}(\mathbb{R}^{n})=M_{0}%
^{1}(\mathbb{R}^{n})$ is the smallest algebra having this property.
\end{proposition}

\noindent(See \cite{gr01}, Theorem 12.1.9, for a proof).

The Feichtinger algebra $S_{0}(\mathbb{R}^{n})$ contains non-differentiable
functions, such as
\[
\psi(x)=\left\{
\begin{array}
[c]{c}%
1-|x|\text{ \textit{if} }|x|\leq1\\
0\text{ \textit{if} }|x|>1
\end{array}
\right.
\]
and it is thus a more general space than the Schwartz space $\mathcal{S}%
(\mathbb{R}^{n})$. This property, together with the fact that Banach spaces
are mathematically easier to deal with than Fr\'{e}chet spaces, makes the
Feichtinger algebra into a tool of choice for the study of wavepackets.

\subsection{Applications to deformation quantization\label{subsapp}}

\subsubsection{The $\ast$-algebra property for the Moyal product}

Comparing formulae (\ref{star}) and (\ref{abtw}) we see that the twisted
product is just the Moyal product with $\hslash=1/2\pi$:
\begin{equation}
A\#B=A\star_{1/2\pi}B. \label{abh}%
\end{equation}
It turns out that more generally $A\star_{\hbar}B$ and $A\#B$ are related in a
very simple way, and this has the following interesting consequence:

\begin{center}
\textit{If} $A,B\in M_{s}^{\infty,1}(\mathbb{R}^{2n})$ \textit{then}
$A\star_{\hbar}B\in M_{s}^{\infty,1}(\mathbb{R}^{2n}).$
\end{center}

More precisely:

\begin{proposition}
\label{propstar}(i) The symbol class $M_{s}^{\infty,1}(\mathbb{R}^{2n})$ is a
Banach $\ast$-algebra with respect to the Moyal product $\star_{\hbar}$ and
the involution $A\longmapsto\overline{A}$: if $A$ and $B$ are in
$M_{s}^{\infty,1}(\mathbb{R}^{2n})$ then $A\star_{\hbar}B$ is also in
$M_{s}^{\infty,1}(\mathbb{R}^{2n})$. (ii) Let $A\in M_{s}^{\infty
,1}(\mathbb{R}^{2n})$ and assume that $A\star_{\hbar}B=I$. Then $B\in
M_{s}^{\infty,1}(\mathbb{R}^{2n})$.
\end{proposition}

\begin{proof}
(i) Using the representation (\ref{abrep}) of the Moyal product one sees
immediately that%
\begin{equation}
(A\star_{\hbar}B)_{\sqrt{\hbar}}=(A_{\sqrt{\hbar}})\#(B_{\sqrt{\hbar}})
\label{ach}%
\end{equation}
where $A_{\sqrt{\hbar}}(z)=A(z\sqrt{\hbar})$, etc. Since $M_{s}^{\infty
,1}(\mathbb{R}^{2n})$ is a Banach $\ast$-algebra for the twisted convolution
$\#$ it thus suffices to prove the equivalence
\begin{equation}
A_{\lambda}\in M_{s}^{\infty,1}(\mathbb{R}^{2n})\Longleftrightarrow A\in
M_{s}^{\infty,1}(\mathbb{R}^{2n}) \label{alequi}%
\end{equation}
for every $\lambda>0$. In fact, since $(A_{\lambda})_{1/\lambda}=A$ it
suffices to show that if $A\in M_{s}^{\infty,1}(\mathbb{R}^{2n})$ then
$A_{\lambda}\in M_{s}^{\infty,1}(\mathbb{R}^{2n})$. Recall that $A\in
M_{s}^{\infty,1}(\mathbb{R}^{2n})$ means that for one (and hence every)
$\Phi\in\mathcal{S}(\mathbb{R}^{2n})$ we have%
\[
||A||_{M_{s}^{\infty,1}}^{\Phi}=\int_{\mathbb{R}^{2n}}\sup_{z}\left[
|V_{\Phi}A(z,\zeta)|v_{s}(z,\zeta)\right]  \mathrm{d}\zeta<\infty
\]
where $V_{\Phi}$ is the short-time Fourier transform defined by
\[
V_{\Phi}A(z,\zeta)=\int_{\mathbb{R}^{2n}}e^{-2\pi i\zeta\cdot z^{\prime}%
}A(z^{\prime})\overline{\Phi(z^{\prime}-z)}\mathrm{d}z^{\prime}.
\]
Performing the change of variables $z^{\prime}\longmapsto\lambda z^{\prime}$
in the formula above we get%
\[
V_{\Phi}A_{\lambda}(z,)=\lambda^{-2n}V_{\Phi_{1/\lambda}}A_{\lambda}(\lambda
z,\lambda^{-1}\zeta)
\]
and hence%
\[
\sup_{z}|V_{\Phi}A_{\lambda}(z,\zeta)|=\lambda^{-2n}\sup_{z}|V_{\Phi
_{\lambda^{-1}}}A(z,\lambda^{-1}\zeta)|
\]
so that%
\begin{align*}
||A_{\lambda}||_{M_{s}^{\infty,1}}^{\Phi}  &  =\lambda^{-2n}\int
_{\mathbb{R}^{2n}}\sup_{z}|V_{\Phi_{1/\lambda}}A(z,\lambda^{-1}\zeta
)|v_{s}(z,\zeta)\mathrm{d}z\mathrm{d}\zeta\\
&  =\int_{\mathbb{R}^{2n}}\sup_{z}|V_{\Phi_{1/\lambda}}A(z,\zeta
)|v_{s}(z,\lambda\zeta)\mathrm{d}z\mathrm{d}\zeta\\
&  \leq\max(1,\lambda^{2s})||A||_{M_{s}^{\infty,1}}^{\Phi_{1/\lambda}}%
\end{align*}
where we have used the trivial inequality $v_{s}(z,\lambda\zeta)\leq
\max(1,\lambda^{2s})$; it follows that $A_{\lambda}\in M_{s}^{\infty
,1}(\mathbb{R}^{2n})$ if $A\in M_{s}^{\infty,1}(\mathbb{R}^{2n})$ which we set
out to prove. Property (ii) follows from Proposition \ref{propsjogr}.
\end{proof}

As a consequence we get the Wiener property for the Moyal product:

\begin{corollary}
Let $A\in M^{\infty,1}(\mathbb{R}^{2n})$ (the Sj\"{o}strand class). If there
exists $B$ such that $A\star_{\hbar}B=I$ then $A\in M^{\infty,1}%
(\mathbb{R}^{2n})$.
\end{corollary}

\begin{proof}
It immediately follows from Proposition \ref{propstar} above using the Wiener
property of the twisted convolution (Proposition \ref{propsjo}(iii)).
\end{proof}

\subsubsection{Regularity results for the star product}

The following result combines the properties of the spaces $M_{s}^{\infty
,1}(\mathbb{R}^{2n})$, viewed as symbol classes, with those of $M_{s}%
^{q}(\mathbb{R}^{n})$.

\begin{proposition}
\label{bounded}Let $A\in M_{s}^{\infty,1}(\mathbb{R}^{2n})$. The operator
$\widehat{A}^{\hbar}=A^{w}(x,-i\hbar\partial_{x})$ is bounded on $M_{s}%
^{q}(\mathbb{R}^{n})$ for every $q$. In fact, there exists a constant $C>0$
independent of $q$ such that following uniform estimate holds%
\[
||\widehat{A}^{\hbar}||_{M_{s}^{q}\longrightarrow M_{s}^{q}}\leq
C||A||_{M_{s}^{\infty,1}}%
\]
for all $A\in M_{s}^{\infty,1}(\mathbb{R}^{2n})$.
\end{proposition}

\begin{proof}
The result is proven for $\hbar=1/2\pi$ in \cite{gr01}, p.320 and p.323. Let
us show that it holds for arbitrary $\hbar$. Noting that $A^{w}(x,-i\hbar
\partial_{x})=B^{w}(x,-i\partial_{x})$ where $B(x,p)=A(x,2\pi\hbar p)$ it
suffices to show that if $A\in M_{s}^{\infty,1}(\mathbb{R}^{2n})$ then $B\in
M_{s}^{\infty,1}(\mathbb{R}^{2n})$. But this follows from Proposition
\ref{proplinv} with the choice%
\[
M=%
\begin{pmatrix}
I & 0\\
0 & 2\pi\hbar I
\end{pmatrix}
\]
for the change of variable.
\end{proof}

Notice that if we take $q=2$, $s=0$ we have $M_{0}^{2}(\mathbb{R}^{n}%
)=L^{2}(\mathbb{R}^{n})$ hence operators with Weyl symbols in $M_{s}%
^{\infty,1}(\mathbb{R}^{2n})$ are bounded on $L^{2}(\mathbb{R}^{n})$; in
particular, using the inclusion (\ref{cm}), we recover the
Calder\'{o}n--Vaillancourt theorem \cite{cv}.

We begin by making the following remark: there are elements of $L_{s}%
^{q}(\mathbb{R}^{2n})$ which do not belong to the range of any wave-packet
transform $W_{\phi}$ (or, equivalently, to the range of any short-time Fourier
transform $V_{\phi}\psi$). This is actually a somewhat hidden consequence of
the uncertainty principle. Choose in fact a measurable function $\Psi$ such
that
\[
\Psi(z)\leq Ce^{-\frac{1}{\hbar}Mz\cdot z}%
\]
where $C>0$ and $M$ is a real symmetric positive-definite matrix; clearly
$\Psi\in L_{s}^{q}(\mathbb{R}^{2n})$, but the existence of $\phi$ and $\psi$
such that $W_{\phi}\psi=\Psi$ is only possible if the matrix $M$ satisfies the
following very stringent condition (see \cite{golu08-1,goluahp}; also
\cite{grzi01}):

\begin{quotation}
\textit{The moduli of the eigenvalues of }$JM$ are all $\leq1$
\end{quotation}

\noindent which is equivalent to the geometric condition:

\begin{quotation}
\textit{The section of the ellipsoid }$Mz\cdot z\leq\hbar$\textit{ by any
plane of conjugate coordinates }$x_{j},p_{j}$\textit{ is }$\geq\pi\hbar.$
\end{quotation}

The properties above are proven by using Hardy's uncertainty principle (Hardy
\cite{ha33}) which is a precise statement of the fact that a function and its
Fourier transform cannot be simultaneously sharply localized. In the
multi-dimensional case this principle can be stated as follows (de Gosson and
Luef \cite{golupre}): if $A$ and $B$ are two real positive definite matrices
and $\psi\in L^{2}(\mathbb{R}^{n})$, $\psi\neq0$ such that
\begin{equation}
|\psi(x)|\leq C_{A}e^{-\tfrac{1}{2\hbar}Ax^{2}}\text{ \ and \ }|F\psi(p)|\leq
C_{B}e^{-\tfrac{1}{2\hbar}Bp^{2}} \label{AB}%
\end{equation}
for some constants $C_{A},C_{B}>0$, then the eigenvalues $\lambda_{j}$,
$j=1,...,n$, of $AB$ are $\leq1$. The statements above then follow, performing
a symplectic diagonalization of $M$ and using the marginal properties of the
cross-Wigner transform.

We will call a function $\Psi\in L_{s}^{q}(\mathbb{R}^{2n})$
\textit{admissible} if there exist $\psi\in M_{s}^{q}(\mathbb{R}^{n})$ and a
window $\phi$ such that $\Psi=W_{\phi}\psi$. Intuitively, the fact for a
function to be admissible means that it is not \textquotedblleft too
concentrated\textquotedblright\ around a phase-space point.

The modulation spaces $M_{s}^{q}(\mathbb{R}^{n})$ can be used to prove the
following regularity result in deformation quantization:

\begin{proposition}
Assume that $A\in M_{s}^{\infty,1}(\mathbb{R}^{2n})$ and that $B\in L_{s}%
^{q}(\mathbb{R}^{2n})$ is admissible. Then $A\star_{\hbar}B\in L_{s}%
^{q}(\mathbb{R}^{2n})$.
\end{proposition}

\begin{proof}
We have
\[
A\star_{\hbar}B=\widetilde{A}^{\hbar}(B)=\widetilde{A}^{\hbar}(W_{\phi}\psi)
\]
for some $\psi\in M_{s}^{q}(\mathbb{R}^{n})$ and a window $\phi$, and hence,
using the first intertwining formula (\ref{fundstar}),%
\[
A\star_{\hbar}B=W_{\phi}(\widehat{A}^{\hbar}\psi).
\]
Since $\psi\in M_{s}^{q}(\mathbb{R}^{n})$ we have $W_{\phi}\psi\in L_{s}%
^{q}(\mathbb{R}^{2n})$ and Proposition \ref{bounded} implies that $\widehat
{A}^{\hbar}\psi\in M_{s}^{q}(\mathbb{R}^{n})$ hence $W_{\phi}(\widehat
{A}^{\hbar}\psi)\in L_{s}^{q}(\mathbb{R}^{2n})$ which we set out to prove.
\end{proof}

\subsubsection{The star-exponential\label{secapp}}

Let $H$ be a Hamiltonian function. In deformation quantization one defines the
star-exponential $\operatorname{Exp}(Ht)$by the formal series
\[
\operatorname{Exp}(Ht)=\sum_{k=0}^{\infty}\frac{1}{k!}\left(  \frac{t}{i\hbar
}\right)  ^{k}(H\star_{\hbar})^{k}%
\]
where $(H\star_{\hbar})^{0}=\operatorname*{Id}$ and $(H\star_{\hbar}%
)^{k}=H\star_{\hbar}(H\star_{\hbar})^{k-1}$ for $k\geq1$. In terms of the Bopp
pseudo-differential operator $\widetilde{H}$ we thus have%
\begin{equation}
\operatorname{Exp}(Ht)=\sum_{k=0}^{\infty}\frac{1}{k!}\left(  \frac{t}{i\hbar
}\right)  ^{k}\widetilde{H}^{k}; \label{a}%
\end{equation}
this formula allows us to \textit{redefine} the star-exponential
$\operatorname{Exp}(Ht)$ by
\begin{equation}
\operatorname{Exp}(Ht)=\exp\left(  -\frac{it}{\hbar}\widetilde{H}\right)  .
\label{redef}%
\end{equation}
With this redefinition $\operatorname{Exp}(Ht)$ is the evolution operator for
the phase-space Schr\"{o}dinger equation%
\begin{equation}
i\hbar\frac{\partial\Psi}{\partial t}=\widetilde{H}\Psi\text{ \ , \ }%
\Psi(\cdot,0)=\Psi_{0}. \label{pse}%
\end{equation}
That is, the solution $\Psi$ of the Cauchy problem (\ref{pse}) is given by%
\begin{equation}
\Psi(z,t)=\operatorname{Exp}(Ht)\Psi_{0}(z). \label{pepsi}%
\end{equation}

Let now
\begin{equation}
U_{t}=\exp\left(  -\frac{it}{\hbar}\widehat{H}^{\hbar}\right)  \label{evo}%
\end{equation}
be the evolution operator for the Schr\"{o}dinger equation%
\begin{equation}
i\hbar\frac{\partial}{\partial t}\psi(x,t)=\widehat{H}^{\hbar}\psi(x,t)\text{
\ , \ }\psi(x,0)=\psi_{0}(x) \label{seq}%
\end{equation}
with Hamiltonian operator $\widehat{H}$. (We will always assume that the
solutions of (\ref{seq}) exist for all $t$ and are unique for an initial datum
$\psi_{0}\in\mathcal{S}(\mathbb{R}^{n})$.)

The following intertwining and conjugation relations are obvious:%
\begin{equation}
\operatorname{Exp}(Ht)W_{\phi}=W_{\phi}U_{t} \label{i2}%
\end{equation}%
\begin{equation}
W_{\phi}^{\ast}\operatorname{Exp}(Ht)=\exp U_{t}W_{\phi}^{\ast} \label{commu}%
\end{equation}%
\begin{equation}
W_{\phi}^{\ast}\operatorname{Exp}(Ht)W_{\phi}=\exp U_{t}. \label{ter}%
\end{equation}

We also note that it immediately follows from formula (\ref{coboppp}) in
Proposition \ref{sycobo} that we have the symplectic covariance formula%

\[
\operatorname{Exp}\left[  (H\circ S^{-1})t\right]  =U_{S}\operatorname{Exp}%
(Ht)U_{S}^{-1}%
\]
where $U_{S}\in\operatorname*{Mp}(4n,\mathbb{R})$ is defined by
\[
U_{S}\Psi(z)=\Psi(Sz)
\]
for $S\in\operatorname*{Sp}(2n,\mathbb{R})$.

The following result shows that the star-exponential preserves the admissible
functions in the weighted $L^{q}$ spaces:

\begin{proposition}
Assume that the Hamiltonian is of the type%
\begin{equation}
H(z)=\frac{1}{2}Mz\cdot z+m\cdot z \label{hm}%
\end{equation}
where $M$ is symmetric and $m\in\mathbb{R}^{2n}$. Let $\Psi\in L_{s}%
^{q}(\mathbb{R}^{2n})$ be admissible. Then
\begin{equation}
\operatorname{Exp}(Ht)\Psi\in L_{s}^{q}(\mathbb{R}^{2n})\text{ \ for all }%
t\in\mathbb{R} \label{utl}%
\end{equation}
for all $q\geq1$ and $s\geq0$.
\end{proposition}

\begin{proof}
Assume first that $m=0$; then the Hamiltonian flow determined by $H$ consists
of symplectic matrices and is thus a one-parameter subgroup $(S_{t})$ of
$\operatorname*{Sp}(2n,\mathbb{R})$. To $(S_{t})$ corresponds a unique
one-parameter subgroup $(\widehat{S}_{t}^{\hbar})$ of the metaplectic group
$Mp(2n,\mathbb{R})$, and we have $U_{t}=\widehat{S}_{t}^{\hbar}$, that is, the
function $\psi(x,t)=\widehat{S}_{t}^{\hbar}\psi_{0}(x)$ is the solution of
Schr\"{o}dinger's equation (\ref{seq}) (see for instance \cite{Birk}, Chapter
7, \S 7.2.2). In view of Proposition \ref{propinvariance}(ii) we have
$\widehat{S}_{t}^{\hbar}:M_{s}^{q}(\mathbb{R}^{n})\longrightarrow M_{s}%
^{q}(\mathbb{R}^{n})$. If $\Psi$ is admissible there exists $\psi\in M_{s}%
^{q}(\mathbb{R}^{n})$ and a window $\phi$ such that $\Psi=W_{\phi}\psi$ hence,
taking formula (\ref{i2}) into account,%
\[
\operatorname{Exp}(Ht)\Psi=W_{\phi}U_{t}\psi;
\]
since $U_{t}\psi\in M_{s}^{q}(\mathbb{R}^{n})$ we have $W_{\phi}U_{t}\psi\in
L_{s}^{q}(\mathbb{R}^{2n})$ hence (\ref{utl}) when $m=0.$ The case $m\neq0$
follows since the one-parameter subgroup $(S_{t})$ of $\operatorname*{Sp}%
(2n,\mathbb{R})$ is replaced by a one-parameter subgroup of the inhomogeneous
(=affine) symplectic group $\operatorname*{ISp}(2n,\mathbb{R})$, from which
follows that $U_{t}=\widehat{S}_{t}^{\hbar}$ is replaced by $U_{t}=\widehat
{S}_{t}^{\hbar}\widehat{T}^{\hbar}(z_{0})$ for some $z_{0}\in\mathbb{R}^{2n}$
only depending on $m$ (see Littlejohn \cite{Littlejohn}); one concludes
exactly as above using the invariance of $M_{s}^{q}(\mathbb{R}^{n})$ under the
action of Weyl--Heisenberg operators (Proposition \ref{propinvariance}(i)).
\end{proof}

\section{Concluding Remarks}

Our results are not the most general possible. The modulation spaces
$M_{s}^{\infty,1}$ and $M_{s}^{q}$ considered in this paper are particular
cases of the more general spaces $M_{m}^{q,r}$ where $q,r$ are real numbers or
$\infty$ and $m$ a more general weight function than $v_{s}$. Our choice was
dictated by the fact that while many of the results we have stated still
remain valid for these more general modulation spaces if certain natural
assumptions (for instance sub-additivity) are made on the weight $m$ the
notation can sometimes appear as too complicated. Another topic we only
briefly mentioned, is the Feichtinger algebra $M_{0}^{1}(\mathbb{R}^{n}%
)=S_{0}(\mathbb{R}^{n})$. In addition to the properties we listed, it has the
following nice feature: let $S_{0}^{\prime}(\mathbb{R}^{n})$ be the dual of
$S_{0}(\mathbb{R}^{n})$; then $(S_{0}(\mathbb{R}^{n}),L^{2}(\mathbb{R}%
^{n}),S_{0}^{\prime}(\mathbb{R}^{n}))$ is a Gelfand triple of Banach spaces;
this property makes $S_{0}(\mathbb{R}^{n})$ particularly adequate for the
study of the continuous spectrum of operators. In addition to smooth
wavepackets (for instance Gaussians).

Another direction certainly worth to be explored is the theory of Wiener
amalgam spaces \cite{fei81,gr01}, which are closely related to modulation
spaces; Cordero and Nicola \cite{coni07} have obtained very interesting
results for the Schr\"{o}dinger equation using Wiener amalgam spaces. What
role do they play in deformation quantization?

\begin{acknowledgement}
Part of this paper was written during a stay of the first author (MdG) at the
Max-Planck-Institut f\"{u}r Mathematik in Bonn in December 2008; he has also
been financed by the Austrian Research Agency FWF (Projektnummer P20442-N13).
The second author (FL) has been supported by the Marie Curie Outgoing
Fellowship PIOF 220464.
\end{acknowledgement}

\end{document}